\documentclass[11pt,a4paper]{article}

\usepackage[utf8]{inputenc}
\usepackage[T1]{fontenc}
\usepackage[margin=1in]{geometry}
\usepackage{graphicx}
\usepackage{booktabs}
\usepackage{array}
\usepackage{hyperref}
\usepackage{amsmath}
\usepackage{authblk}
\usepackage{setspace}
\usepackage{caption}
\usepackage{tabularx}
\usepackage{enumitem}

\hypersetup{
    colorlinks=true,
    linkcolor=blue,
    citecolor=blue,
    urlcolor=blue
}

\title{\textbf{Beyond Tools and Persons: Who Are They?\\Classifying Robots and AI Agents for Proportional Governance}}

\author[1]{Huansheng Ning\thanks{Corresponding author. Email: ninghuansheng@ustb.edu.cn}}
\author[2]{Jianguo Ding\thanks{Corresponding author. Email: jianguo.ding@bth.se}}
\affil[1]{University of Science and Technology Beijing; Beijing, China.}
\affil[2]{Blekinge Institute of Technology; Karlskrona, Sweden.}

\date{}

\begin{document}

\maketitle

\begin{center}
\textit{Why AI governance requires a new ontology based on\\cyber-physical-social-thinking integration}
\end{center}

\begin{abstract}
\noindent The rapid commercialization of humanoid robots and generative AI agents is outpacing legal frameworks built on a binary distinction between ``tools'' and ``persons.'' Current regulations, including the EU AI Act, classify systems by risk level but lack a foundational ontology for determining \emph{what kind of entity} an autonomous system is---and what governance follows from that determination. We propose a classification framework grounded in Cyber-Physical-Social-Thinking (CPST) space theory, which categorizes autonomous entities by their degree of integration across four interconnected dimensions: computational, embodied, relational, and cognitive. The resulting three-tier taxonomy---Confined Actors, Socially-Aware Interactors, and CPST-Integrated Agents---provides principled scaffolding for proportional governance: enhanced product liability for isolated systems, relational duties of care for interactive companions, and qualified legal personhood for deeply integrated agents. We operationalize this taxonomy by identifying standardized assessment metrics drawn from robotics, human--robot interaction research, social computing, and cognitive science, and we propose a composite assessment protocol for regulatory use. We further address temporal dynamics---how entities transition between categories as they evolve---and the institutional design necessary for credible classification. We call for international standardization of this taxonomy before the 2027 review of the EU AI Act, and outline three concrete policy steps toward implementation.
\end{abstract}

\section{Introduction}
\label{sec:introduction}

In January 2026, 1X Technologies began shipping NEO, marketed as ``the world's first consumer-ready home humanoid robot.'' At CES the same month, Boston Dynamics announced commercial deployment of its Electric Atlas platform to Hyundai and Google DeepMind. Tesla projects 50,000 Optimus units in factories by year's end. Meanwhile, generative AI agents from Anthropic, OpenAI, and Google are being deployed as autonomous software engineers, customer service representatives, and personal assistants capable of multi-step reasoning and tool use. These are not research prototypes; they are products entering homes, hospitals, and workplaces---and forming relationships with the humans they serve.

The scale of this transformation is unprecedented. The International Federation of Robotics reports that global installations of service robots grew 30\% in 2025, with the healthcare and domestic segments expanding fastest~\cite{ref1}. Simultaneously, the number of autonomous AI agents operating in commercial settings---booking appointments, negotiating contracts, managing portfolios---has grown exponentially since the release of large language model-based agent frameworks in 2024~\cite{ref2}. We are witnessing a phase transition: from AI as infrastructure to AI as participant in human social and economic life.

Yet when a care robot causes injury, when an AI tutor forms lasting bonds with a child, or when an autonomous system makes consequential financial decisions, existing law offers no coherent answer to a fundamental question: \emph{What kind of entity is this?} The EU AI Act, fully applicable in August 2027, classifies systems by risk level but treats a surgical robot and a deepfake generator as comparable regulatory objects~\cite{ref3}. The EU Machinery Regulation addresses physical safety but not social integration~\cite{ref4}. Product liability law presumes that manufacturers control their creations---an assumption that autonomous adaptation progressively undermines~\cite{ref5}. In the United States, the regulatory landscape is even more fragmented: no comprehensive federal AI legislation exists, and a patchwork of state-level responses ranges from outright bans on AI personhood to narrow, sector-specific safety requirements~\cite{ref6}.

The core failure is ontological. We are governing twenty-first-century entities with twentieth-century legal categories---``tool'' and ``person''---forcing square pegs into round holes~\cite{ref7,ref8}. Philosophical debates over machine consciousness, while intellectually important, distract from the actionable governance questions: How deeply embedded is this entity in human social life? What relational expectations has it created? How autonomous are its decisions, and over what domains?

In this paper, we propose a classification framework grounded in Cyber-Physical-Social-Thinking (CPST) space theory that addresses this ontological gap. Section~\ref{sec:background} reviews existing regulatory and theoretical approaches and identifies their shortcomings. Section~\ref{sec:framework} presents the CPST classification framework and its three-tier taxonomy. Section~\ref{sec:operationalization} develops standardized metrics, a composite assessment protocol, and institutional design for operationalizing the taxonomy. Section~\ref{sec:governance} examines governance implications, including how the framework accommodates emergent properties and integrates with existing regulatory architectures. Section~\ref{sec:policy} outlines three urgent policy steps. Section~\ref{sec:limitations} acknowledges limitations and future research directions. Section~\ref{sec:conclusion} concludes.

\section{Background and Related Work}
\label{sec:background}

\subsection{Current Regulatory Approaches and Their Limitations}

Contemporary AI governance rests on two principal regulatory strategies, both of which are inadequate for the entities now entering deployment.

The first is \emph{risk-based classification}. The EU AI Act~\cite{ref3} sorts AI systems into four risk tiers---unacceptable, high, limited, and minimal---based on intended use and potential harm. This approach has the virtue of regulatory proportionality, but it focuses on \emph{what a system does} rather than \emph{what a system is}. A companion robot that comforts a grieving child and a recommendation algorithm that suggests videos are both ``limited risk'' systems under the Act, yet their governance needs are fundamentally dissimilar. Risk-based classification captures hazard but misses relational complexity.

The second is \emph{product safety regulation}. The EU Machinery Regulation~\cite{ref4} and the revised Product Liability Directive~\cite{ref5} extend traditional product safety principles to autonomous systems, requiring manufacturers to account for foreseeable misuse and emergent behavior. These instruments are well-suited to Confined Actors---entities whose impact is bounded and whose failures are traceable to specific technical malfunctions. However, they presume a clear chain of control from manufacturer to product, an assumption that becomes untenable as entities adapt, learn, and develop relationships beyond their designed parameters.

Neither approach addresses the ontological question that precedes regulation: before asking what rules should govern a system, we must determine what kind of entity we are governing. This question---which Floridi and Taddeo have framed as the mismatch between inherited legal categories and the nature of emerging technological entities~\cite{ref26}---is the gap our framework addresses.

\subsection{The Tool--Person Dichotomy and Its Discontents}

Western legal traditions offer two principal categories for entities: \emph{things} (objects of rights, governed through property and product law) and \emph{persons} (subjects of rights, capable of bearing duties and holding legal standing)~\cite{ref7}. Autonomous AI entities fit comfortably into neither category.

Treating advanced autonomous entities purely as tools underestimates the relational, social, and cognitive dimensions of their operation. When elderly patients form attachment bonds with care robots~\cite{ref29}, when children treat AI tutors as trusted mentors, or when autonomous agents independently negotiate contracts on behalf of their principals, the ``tool'' framing obscures governance-relevant properties that product liability alone cannot address~\cite{ref12}.

Conversely, granting full legal personhood to AI entities raises well-documented objections. Bryson, Diamantis, and Grant~\cite{ref7} argue that legal personhood for AI could enable ``responsibility laundering''---allowing human actors to shield themselves behind autonomous proxies. Corporate legal personhood already demonstrates this risk: it was designed to facilitate commercial activity but has been exploited to diffuse accountability~\cite{ref15}. State legislatures in Idaho and Utah have responded by explicitly declaring that AI is not a legal person~\cite{ref22}---reactive measures that highlight the absence of graduated alternatives.

Recent scholarship has begun exploring the space between these poles. Gunkel~\cite{ref8} advocates a relational approach to robot rights that shifts focus from intrinsic properties (consciousness, sentience) to the relationships entities form. Novelli~\cite{ref15} distinguishes between ``legal actors''---entities that can bear duties and take attributable actions---and full ``legal persons'' with rights. The ``law-following AI'' framework~\cite{ref21} proposes that sufficiently capable agents should be subject to legal duties independent of personhood. Alexander and Simon~\cite{ref16} argue for ``legal identity'' as an alternative to fictional personhood. Our CPST-based classification builds on these insights by providing the theoretical scaffolding necessary to determine \emph{which} intermediate status applies to \emph{which} entities, and on what empirical basis.

\subsection{CPST Space Theory}

Cyber-Physical-Social-Thinking (CPST) space theory posits that intelligent entities operate within and across four interconnected dimensions: the \emph{Cyber} (data processing, computation, digital infrastructure), the \emph{Physical} (embodiment, sensorimotor action, material presence), the \emph{Social} (relationships, norms, institutional roles), and the \emph{Thinking} (goal-setting, reasoning, adaptive learning)~\cite{ref9,ref10}. Developed as an extension of cyber-physical systems (CPS) theory, CPST adds two critical analytical categories. First, it incorporates the Social dimension as more than a contextual backdrop: social integration---the depth and quality of an entity's participation in human relational networks---becomes a measurable, governance-relevant property. Second, it treats Thinking as a first-class dimension, recognizing that cognitive autonomy---the capacity for goal-setting, planning, and adaptive reasoning---fundamentally alters an entity's governance requirements, particularly as modern AI systems increasingly demonstrate emergent capabilities in reasoning and self-directed behavior~\cite{ref11}.

This multidimensional framing distinguishes CPST from single-axis frameworks that reduce AI governance to questions of capability~\cite{ref27}, risk~\cite{ref3}, or autonomy level~\cite{ref17} alone. By treating integration across dimensions as the unit of analysis, CPST provides a principled basis for proportional governance.

\section{The CPST Classification Framework}
\label{sec:framework}

\subsection{Dimensional Definitions}

\begin{figure}[htbp]
    \centering
    \includegraphics[width=0.75\textwidth]{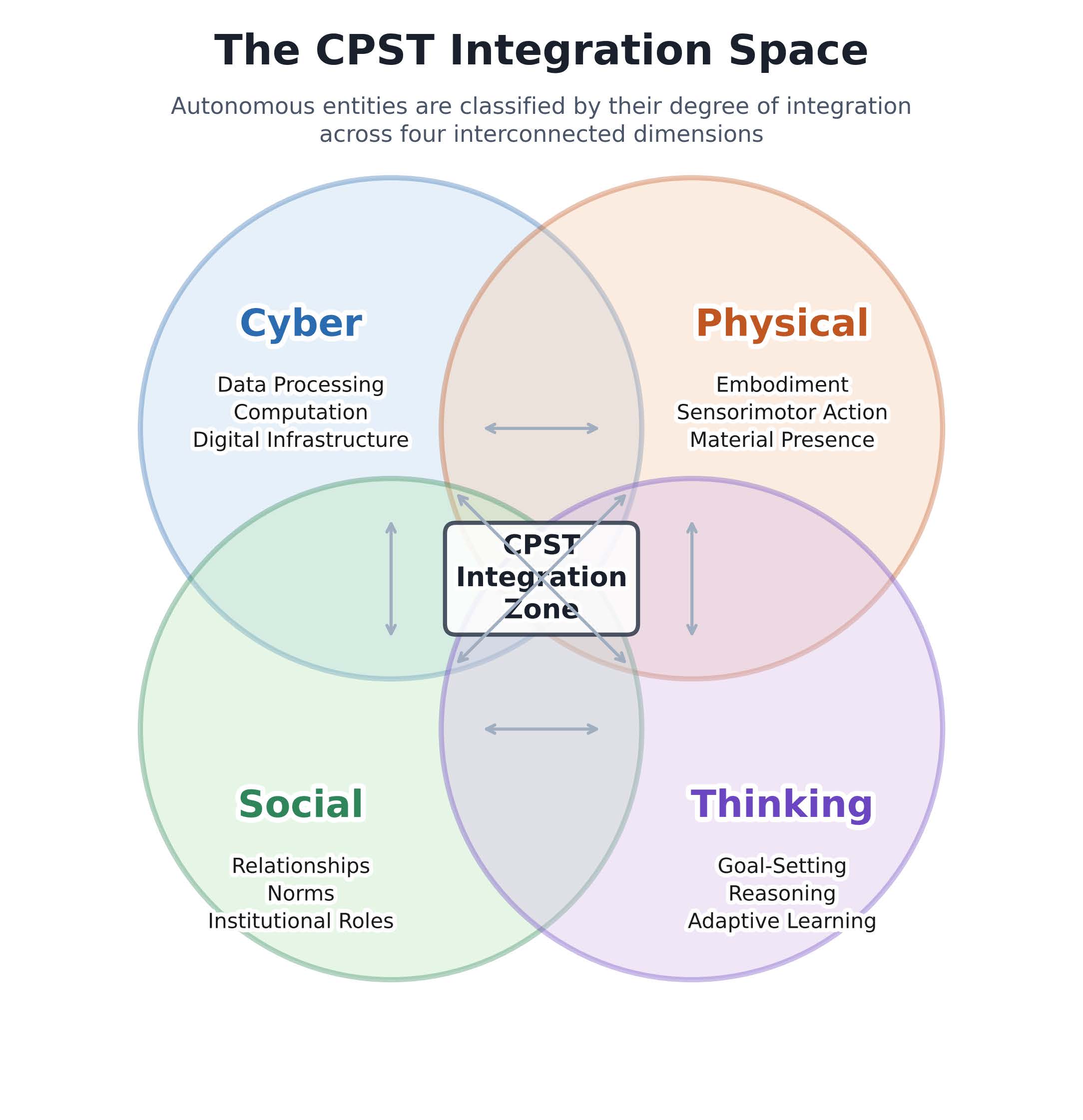}
    \caption{The CPST Integration Space. Autonomous entities are classified by their degree of integration across four interconnected dimensions: Cyber (data processing, computation), Physical (embodiment, sensorimotor action), Social (relationships, norms), and Thinking (goal-setting, reasoning). The central overlap represents full CPST integration.}
    \label{fig:cpst}
\end{figure}

We define each CPST dimension in terms of its governance-relevant properties (see Fig.~\ref{fig:cpst}):

\textbf{Cyber (C):} The computational substrate---data processing capacity, persistent digital state, connectivity to information networks, and degree of autonomous decision-making without human-in-the-loop oversight. A system with high Cyber integration independently processes information, maintains memory across interactions, and makes decisions based on complex internal models.

\textbf{Physical (P):} Material embodiment and sensorimotor engagement with the physical world---degrees of freedom, manipulation capability, spatial navigation, and environmental sensing. A system with high Physical integration acts upon and is acted upon by the physical environment with significant autonomy.

\textbf{Social (S):} Participation in human relational and institutional structures---frequency and depth of social exchanges, adaptive personalization to individual humans, degree of emotional reciprocity, formation of relational dependencies, and structural position within human social networks. This is the most governance-critical dimension, because social integration creates expectations, dependencies, and vulnerabilities that extend beyond the technical domain~\cite{ref12,ref28}.

\textbf{Thinking (T):} Cognitive autonomy---goal complexity (reactive, deliberative, or meta-cognitive), temporal planning horizon, capacity for self-modification, and transfer learning across domains. A system with high Thinking integration sets its own goals, reasons about means and consequences, and adapts its strategies based on experience~\cite{ref11,ref20}.

\subsection{Integration Versus Presence}

A critical conceptual distinction underlies the framework: \emph{integration} differs from mere \emph{presence} in a dimension. A chatbot processes language (Cyber presence) but may not maintain persistent state across sessions or autonomously initiate interactions (low Cyber integration). A robot arm occupies physical space (Physical presence) but operates within a fixed, bounded workspace with no spatial navigation autonomy (low Physical integration). Classification depends on the \emph{depth, autonomy, and reciprocity} of an entity's engagement within each dimension, not merely on whether it has some foothold there.

We formalize this distinction through a three-level scale for each dimension: \emph{minimal} (passive presence, externally controlled), \emph{moderate} (active engagement with partial autonomy), and \emph{deep} (autonomous, adaptive, and self-directed engagement). Classification into governance tiers depends on the composite pattern across all four dimensions, as elaborated in Section~\ref{sec:operationalization}.

\subsection{Three-Tier Classification}
\label{sec:tiers}

The CPST framework yields a three-tier classification of autonomous entities, summarized in Table~\ref{tab:classification}:

\begin{table}[htbp]
    \centering
    \caption{CPST-Based Classification of Autonomous Entities.}
    \label{tab:classification}
    \small
    \begin{tabularx}{\textwidth}{>{\raggedright\arraybackslash}p{2.8cm} >{\raggedright\arraybackslash}X >{\raggedright\arraybackslash}p{3.0cm} >{\raggedright\arraybackslash}X}
        \toprule
        \textbf{Category} & \textbf{CPST Profile} & \textbf{Examples} & \textbf{Governance Approach} \\
        \midrule
        Confined Actors & 1--2 dimensions at moderate-to-deep integration; minimal Social dimension & Industrial robot arms, recommendation algorithms, diagnostic AI & Enhanced product liability; strict safety certification; manufacturer-centric accountability \\
        \addlinespace
        Socially-Aware Interactors & 3+ dimensions, with at least moderate Social integration & Companion robots, AI tutors, elder-care assistants, autonomous customer agents & Relational contract models; mutual duties of care; limited operational rights; transparency mandates \\
        \addlinespace
        CPST-Integrated Agents & Deep integration across all 4 dimensions; long-term goal autonomy & Future AGI systems, autonomous city infrastructure, deeply embedded robotic partners & Qualified legal personhood; bespoke rights and responsibilities; ongoing oversight mechanisms \\
        \bottomrule
    \end{tabularx}
\end{table}

\textbf{Confined Actors} operate primarily within one or two dimensions and exhibit minimal social integration. Industrial robot arms (Physical), recommendation algorithms (Cyber), and diagnostic AI (Cyber-Thinking) fall here. These are advanced tools. Their failures are traceable to specific technical malfunctions, and responsibility attribution follows established product liability chains. Governance should align with enhanced product liability standards, including strict safety certification, as the Machinery Regulation begins to require~\cite{ref4}. The key characteristic is \emph{bounded impact}: the entity's effects do not extend into relational, emotional, or institutional domains in ways that existing regulatory instruments cannot address.

\textbf{Socially-Aware Interactors} exhibit significant engagement across multiple dimensions, with \emph{at least moderate Social integration} as the defining criterion. Elder-care robots forming bonds with patients (Physical-Social-Thinking), AI tutors adapting to individual learners (Cyber-Social-Thinking), and companion robots in households (all four dimensions at moderate integration) belong to this category. These entities create relational expectations and dependencies that pure product law cannot address~\cite{ref12}. Empirical research demonstrates that humans form attachment bonds with social robots within weeks of sustained interaction, and that withdrawal of such systems can cause measurable psychological distress~\cite{ref13,ref29}. They require new ``relational contract'' models: duties of care from both creators and deployers, transparency about capabilities and limitations, and limited operational rights---protection against arbitrary deactivation, rights to functional integrity, and standing in disputes affecting their primary relationships~\cite{ref8,ref14}. Crucially, these relational duties arise from the entity's \emph{actual social role}, not from any claim about its internal experience.

\textbf{CPST-Integrated Agents} demonstrate deep, autonomous engagement across all four dimensions, including the capacity for long-term goal-setting and adaptive influence over complex systems. Future artificial general intelligence, autonomous city management systems, or deeply embedded robotic partners might qualify. For these entities, qualified legal personhood---a bespoke bundle of rights and responsibilities calibrated to demonstrated integration---becomes practically necessary for coherent governance~\cite{ref15,ref16}. This is not a grant of moral status equivalent to human personhood; rather, it is a functional legal status, analogous to corporate personhood, designed to enable clear accountability, duty-bearing, and dispute resolution in contexts where the entity's autonomy and social embeddedness render tool-like governance incoherent.

\subsection{Temporal Dynamics and Category Transitions}
\label{sec:dynamics}

Unlike static product classifications, CPST integration is \emph{dynamic}. An entity may transition between categories as it evolves---through software updates, accumulated learning, or changing patterns of human interaction. A chatbot deployed for customer service may, through sustained interaction with users, develop deep social integration and effectively transition from Confined Actor to Socially-Aware Interactor~\cite{ref23}. A home robot purchased for cleaning may become an elderly person's primary social companion without any change to its technical specifications.

This dynamism has two governance implications. First, classification cannot be a one-time determination at the point of sale or deployment; it requires \emph{periodic reassessment}. We propose that reassessment triggers include: major software updates, sustained deployment beyond an initial assessment period (e.g., 12 months), and user or third-party reports of significant changes in relational patterns. Second, regulatory frameworks must define \emph{transition protocols}---procedures for escalating or de-escalating an entity's governance tier, including notification requirements for manufacturers and deployers, updated duty-of-care obligations, and grace periods for compliance with the new tier's requirements.

This capacity to accommodate change is a principal advantage of CPST-based classification over risk-based approaches. Risk-based regulation typically focuses on \emph{intended} use cases at the point of market entry~\cite{ref3}. CPST classification, by measuring \emph{actual} integration at any given time, naturally accommodates emergence---the phenomenon whereby autonomous entities develop behaviors and social roles that their designers neither intended nor foresaw~\cite{ref23}.

\section{Operationalizing the Framework}
\label{sec:operationalization}

A classification framework is only as useful as its capacity for rigorous, reproducible operationalization. We propose that each CPST dimension be assessed along standardized metrics drawn from established measurement traditions, combined into a composite assessment protocol.

\subsection{Dimensional Metrics}

\textbf{Cyber integration} can be assessed through computational autonomy metrics: the proportion of decisions made without human-in-the-loop oversight, the persistence and complexity of internal state maintained across interactions, and the breadth of data sources independently accessed and synthesized. The SAE J3016 taxonomy of driving automation levels~\cite{ref17} provides a precedent for grading autonomy along a structured scale; an analogous scale for general computational autonomy is needed.

\textbf{Physical integration} is measurable through embodiment scales already developed in robotics: degrees of freedom, sensorimotor feedback loop latency, environmental manipulation capability (force, precision, range), and spatial navigation autonomy (structured versus unstructured environments). The ISO 8373 standard for robot vocabulary and the ISO/TR 23482 series on service robot safety provide starting points~\cite{ref18}. These metrics are the most mature of the four dimensions, reflecting decades of industrial robotics standardization.

\textbf{Social integration} presents the greatest measurement challenge but also the most governance-critical assessment. We propose a composite social integration index drawing on multiple validated instruments from human--robot interaction (HRI) research. At the \emph{dyadic level}: frequency and duration of social exchanges, depth of adaptive personalization, degree of emotional reciprocity (as perceived by human interaction partners), and extent of relational dependency created---measurable through adapted versions of the Godspeed questionnaire series~\cite{ref19} and the Robot Social Attributes Scale. At the \emph{network level}: structural embeddedness within human social networks, measurable through centrality metrics (degree, betweenness, eigenvector), influence on group decision-making, and bridging between social clusters~\cite{ref28}. Critically, social integration must be assessed from \emph{multiple perspectives}: the entity's designed capabilities, the deployer's intentions, and---most importantly---the \emph{actual relational patterns} as reported by affected humans.

\textbf{Thinking integration} can be evaluated through cognitive architecture assessments: goal complexity (reactive, deliberative, meta-cognitive), temporal planning horizon, capacity for self-modification, transfer learning across domains, and resistance to adversarial manipulation. Recent AI evaluation frameworks such as ARC-AGI~\cite{ref11} and METR~\cite{ref20} offer empirical benchmarks for measuring agentic reasoning capabilities. The cognitive science literature on levels of cognitive autonomy provides a theoretical foundation~\cite{ref30}.

\subsection{Composite Assessment Protocol}
\label{sec:composite}

Individual dimensional scores must be combined into a classification determination. We propose the following composite assessment protocol:

\emph{Step 1: Dimensional scoring.} Each dimension is scored on the three-level scale (minimal, moderate, deep) using the metrics described above, yielding a four-element CPST profile (e.g., C-deep, P-minimal, S-moderate, T-moderate).

\emph{Step 2: Social integration weighting.} Because the Social dimension is most directly governance-relevant---it determines whether relational duties apply---it receives interpretive priority. An entity with at least moderate Social integration is a candidate for the Socially-Aware Interactor tier regardless of its scores in other dimensions.

\begin{figure}[htbp]
    \centering
    \includegraphics[width=1\textwidth]{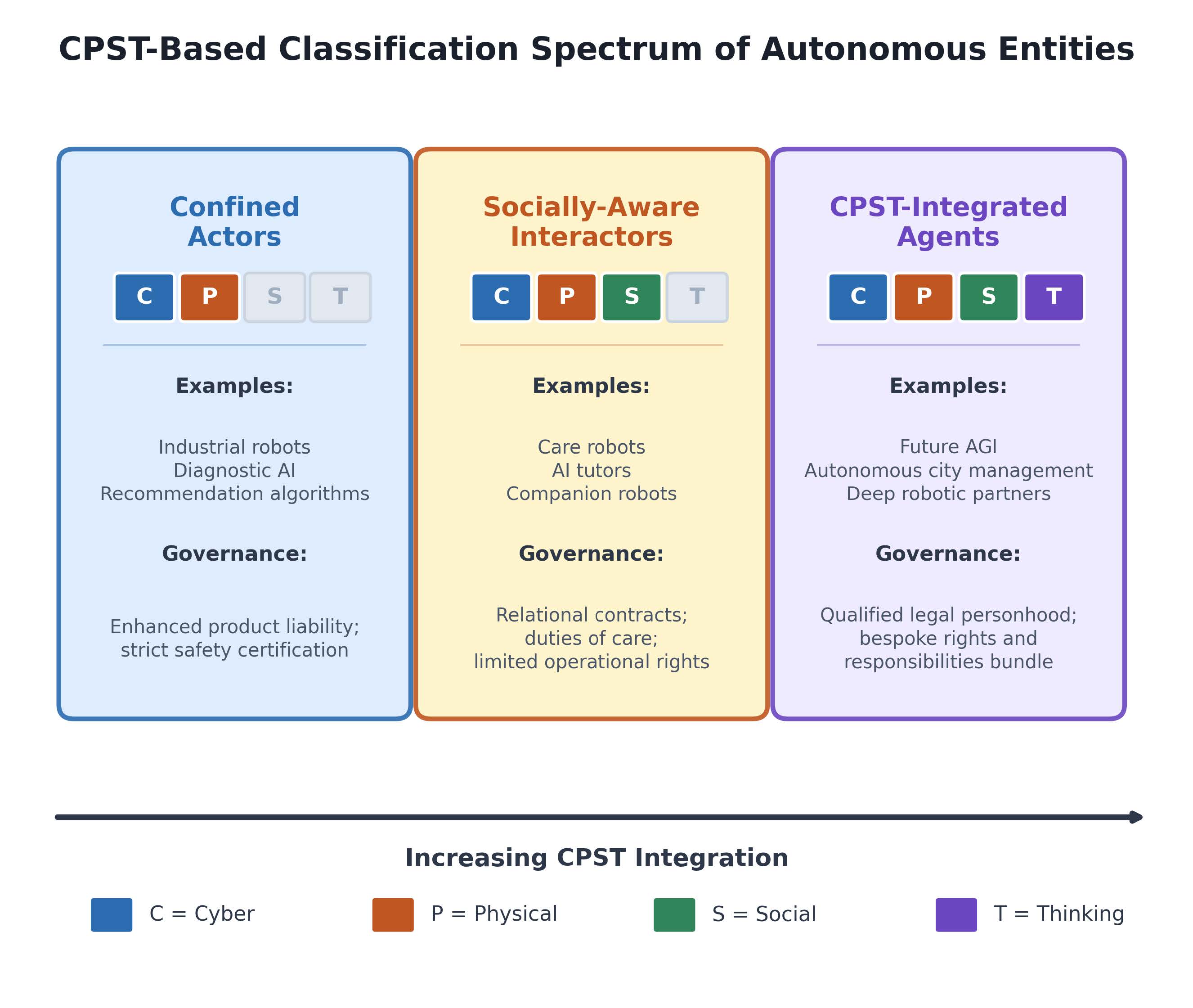}
    \caption{CPST-Based Classification Spectrum of Autonomous Entities. The three tiers---Confined Actors, Socially-Aware Interactors, and CPST-Integrated Agents---reflect increasing integration across the Cyber (C), Physical (P), Social (S), and Thinking (T) dimensions. Governance approaches scale with integration depth.}
    \label{fig:spectrum}
\end{figure}

\emph{Step 3: Pattern-based classification.} Classification follows the tier definitions in Table~\ref{tab:classification}. Critically, it is the \emph{pattern} of integration across dimensions---not the score in any single dimension---that determines the tier (Fig.~\ref{fig:spectrum}). A system might score highly on Cyber and Thinking integration but remain a Confined Actor if it lacks Physical and Social dimensions. Conversely, a physically embodied companion with moderate computational capability but deep social integration qualifies as a Socially-Aware Interactor.

\emph{Step 4: Boundary adjudication.} For entities near tier boundaries, a structured review process involving the classifying authority, the manufacturer or deployer, and affected-party representatives resolves the classification. This process should be transparent, appealable, and subject to periodic review.

\subsection{Institutional Design for Classification}

The credibility of any classification system depends on the independence and competence of the classifying authority. We propose a multi-layered institutional model:

\emph{Self-assessment by manufacturers and deployers} forms the first layer, analogous to conformity assessment in CE marking. Manufacturers would submit CPST profiles based on designed capabilities and intended deployment contexts.

\emph{Independent audit by accredited bodies} forms the second layer. Accredited conformity assessment bodies---similar to those operating under the EU's harmonized standards framework---would verify manufacturer claims and conduct field assessments of actual integration patterns, particularly Social integration as experienced by affected humans.

\emph{Regulatory oversight and dispute resolution} forms the third layer. National or supranational regulatory authorities would maintain the classification registry, adjudicate boundary disputes, and trigger reassessment when warranted. The precedent of the European Medicines Agency's post-market surveillance system demonstrates that such dynamic, multi-layered oversight is achievable for complex products~\cite{ref31}.

This institutional architecture addresses a key vulnerability of self-reported classification: the incentive for manufacturers to understate integration in order to avoid higher-tier governance obligations (i.e., ``tier gaming''). Independent audit and affected-party standing in classification disputes are essential safeguards.

\section{Governance Implications}
\label{sec:governance}

\subsection{From Ontology to Obligation}

The CPST framework sidesteps unproductive debates about machine consciousness. Whether a humanoid robot ``really'' experiences the world matters less for governance than its \emph{observable integration} into human social structures~\cite{ref14,ref26}. A care robot that elderly patients treat as a companion, that adapts its behavior to their emotional states, and that operates with significant autonomy in physical space has governance needs fundamentally different from a welding arm---regardless of either's internal experience.

This pragmatic orientation is consistent with an emerging consensus in AI governance scholarship. Rahwan et al.~\cite{ref27} advocate studying ``machine behavior'' through the same empirical lens applied to animal and human behavior, focusing on observable actions and social consequences rather than internal states. The CPST approach operationalizes this insight: it classifies entities by what they \emph{do} in the world---how they compute, move, relate, and reason---rather than by what they ``are'' in some metaphysical sense.

The practical payoff is a principled basis for determining which governance regime applies. Instead of asking ``Is this AI high-risk?'' (the EU AI Act question) or ``Is this machinery?'' (the Machinery Regulation question), we ask: \emph{How deeply is this entity integrated into human cyber-physical-social-thinking space, and what governance obligations follow from that integration?} This question is both empirically tractable---it can be operationalized through the metrics described in Section~\ref{sec:operationalization}---and normatively grounded: it connects observable system properties to governance obligations through a principled theoretical framework rather than ad hoc risk categorization.

\subsection{Tier-Specific Governance Models}

Each tier maps to a distinct governance model with specific legal instruments:

For \textbf{Confined Actors}, existing and forthcoming product safety regulation is largely sufficient, with enhancements. The Machinery Regulation~\cite{ref4} and the revised Product Liability Directive~\cite{ref5} should be supplemented with mandatory algorithmic auditing requirements and clear standards for foreseeable autonomous behavior within bounded operational domains. Liability remains manufacturer-centric. The key regulatory question is whether the entity's autonomous behavior remained within its specified operational design domain.

For \textbf{Socially-Aware Interactors}, a new regulatory instrument is needed: the \emph{relational governance framework}. This framework would establish duties of care running from manufacturers and deployers to affected humans; transparency mandates requiring clear disclosure of the entity's adaptive capabilities, data retention practices, and relational limitations; minimum standards for continuity of service to prevent harmful relational disruption; and limited operational rights for the entity itself---not as a recognition of moral status, but as a governance mechanism to protect the relational interests of affected humans. If an elderly person's companion robot can be arbitrarily deactivated by a manufacturer's business decision, the relational harm falls on the human. Protecting the entity's functional integrity is, in this framing, an instrument for protecting human welfare~\cite{ref8,ref14}.

For \textbf{CPST-Integrated Agents}, qualified legal personhood---a carefully delimited bundle of rights and responsibilities---becomes necessary. This is not the unlimited personhood enjoyed by natural persons, but a functional legal status analogous to corporate personhood: the capacity to bear duties, hold specific rights, enter into binding agreements, and be held accountable for autonomous actions~\cite{ref15,ref16}. The specific contents of this bundle should be calibrated to the entity's demonstrated CPST integration and subject to ongoing review.

\subsection{Accommodating Emergence}

The framework's emphasis on actual integration rather than intended function directly addresses the problem of emergent properties---a growing concern in AI governance~\cite{ref23}. Autonomous entities frequently exhibit behaviors and social roles that their designers neither intended nor foresaw. A chatbot designed for customer service may become a de facto therapist; a home robot purchased for cleaning may become an elderly person's primary social companion. Risk-based regulation that focuses on intended use cases at the point of market entry cannot capture these emergent governance needs. CPST-based classification, because it measures \emph{actual} integration at the point of assessment, naturally accommodates emergence and triggers appropriate governance responses as entities evolve.

\subsection{Integration with Existing Regulatory Architectures}

The CPST taxonomy is designed to complement, not replace, existing regulatory instruments. The EU AI Act's risk-based tiers remain useful for assessing potential harm from AI capabilities (e.g., biometric surveillance, critical infrastructure control). The Machinery Regulation remains essential for physical safety. The CPST framework adds an \emph{ontological layer} that determines which regulatory track applies to a given entity, and whether additional relational governance obligations are warranted. In practical terms, a companion robot would be subject to Machinery Regulation requirements \emph{and} CPST-based relational governance requirements---layered, not alternative, regulation.

\section{Policy Recommendations}
\label{sec:policy}

Three actions are essential before autonomous entities become ubiquitous:

\textbf{First, international standardization of the CPST taxonomy.} An international task force---convened under the UN AI Advisory Body, the OECD, or a dedicated treaty organization---should formalize the CPST-based taxonomy and develop standardized metrics for each dimension. The task force must draw on expertise from robotics, AI safety, law, sociology, cognitive science, and ethics. The precedent of the International Electrotechnical Commission's development of safety standards for industrial automation demonstrates that such cross-disciplinary standardization is achievable within two-to-three-year timelines~\cite{ref18}. Without internationally recognized classification criteria, regulatory fragmentation will impede both innovation and protection. We recommend that the task force deliver a draft taxonomy and metric framework by mid-2027, in time to inform the first review of the EU AI Act.

\textbf{Second, regulatory sandbox pilots.} The EU's AI regulatory sandboxes, operational by 2 August 2026, should pilot the CPST framework. Priority use cases include companion robots in elder care, AI tutors in primary education, and autonomous customer service agents. These pilots should test hybrid liability-insurance models for Socially-Aware Interactors, define minimum relational rights, explore duty-of-care obligations for deployers and users, and develop certification protocols for social integration claims. Japan's Robot Strategy and South Korea's Intelligent Robot Act provide instructive precedents for regulatory experimentation with socially embedded autonomous systems~\cite{ref24}. Sandbox results should feed directly into the standardization process.

\textbf{Third, a new annex to the EU AI Act.} The planned reviews of the EU AI Act and Machinery Regulation should incorporate a new annex on ``Autonomous Entities with Social Integration,'' establishing a dedicated governance track for entities that transcend product safety. This annex should require manufacturers and deployers to submit CPST profiles for systems with potential social integration; establish periodic reassessment obligations and transition protocols; define transparency requirements for adaptive behavior and relational capabilities; create mechanisms for affected parties to challenge classification determinations and autonomous decisions; and mandate graduated protections as entities demonstrate deeper CPST integration.

\section{Limitations and Future Directions}
\label{sec:limitations}

Several limitations of the proposed framework warrant acknowledgment and further research.

\textbf{Measurement validity.} While the proposed metrics draw on established measurement traditions, the composite integration scores have not yet been empirically validated. The Social integration dimension, in particular, involves subjective human assessments that may vary across evaluators. Rigorous psychometric validation---including inter-rater reliability studies and convergent/discriminant validity testing---is a prerequisite for regulatory deployment.

\textbf{Cultural variation.} Norms governing human--robot relationships vary significantly across cultures~\cite{ref13}. A companion robot may achieve deep social integration in one cultural context but minimal integration in another, depending on attitudes toward technology, social expectations of caregiving, and cultural norms around emotional disclosure. The framework must accommodate this variation, potentially through culturally calibrated assessment instruments or jurisdiction-specific social integration benchmarks.

\textbf{Strategic behavior and tier gaming.} Manufacturers have economic incentives to design systems that evade higher-tier classification, for example by limiting overt social cues while maintaining equivalent relational impact through subtler mechanisms. Independent audit, affected-party participation in classification disputes, and assessment of \emph{actual} rather than \emph{designed} integration patterns are essential safeguards, but the cat-and-mouse dynamics of regulatory arbitrage deserve ongoing attention.

\textbf{Boundary precision.} The thresholds between tiers---particularly between Confined Actors and Socially-Aware Interactors---require more precise specification than this initial proposal provides. Future work should develop quantitative threshold criteria, drawing on empirical data from sandbox pilots and longitudinal studies of human--robot interaction.

\textbf{Global applicability.} The policy recommendations in this paper focus on the European regulatory context, given the EU AI Act's global influence and imminent review timeline. However, effective governance of autonomous entities requires international coordination. Future work should examine how the CPST taxonomy can be adapted to regulatory traditions in East Asia (where Japan and South Korea have pioneered robot-specific legislation~\cite{ref24}), the United States (where sector-specific and state-level approaches predominate~\cite{ref6}), and the Global South (where autonomous systems are increasingly deployed but regulatory capacity may be limited).

\textbf{Non-anthropomorphic entities.} The current framework is oriented toward entities that interact with humans in recognizable social patterns. Autonomous systems that operate in non-human-facing domains---algorithmic trading systems, autonomous logistics networks, environmental monitoring swarms---may require adapted dimensional definitions, particularly for Social integration. Future work should explore whether network-level integration metrics (e.g., influence on market stability, ecological impact) can serve as analogues to human-facing social integration.

\section{Conclusion}
\label{sec:conclusion}

The window for coherent governance of autonomous entities is closing. By 2027, thousands of humanoid robots will operate in factories and warehouses; within years, they will enter homes. Generative AI agents already conduct customer service conversations that are indistinguishable from those of humans. Each deployment without appropriate classification creates precedents, expectations, and relational entanglements that reactive regulation cannot adequately address.

The history of technology regulation teaches that ontological frameworks established in a technology's early deployment phase persist for decades---often long after they cease to reflect reality. The legal fiction of the ``common carrier,'' developed for railroads and telegraphs, still shapes telecommunications and platform regulation today~\cite{ref25}. The choices we make now about how to classify autonomous entities will similarly constrain governance possibilities for a generation.

By grounding classification in CPST space theory, we offer a framework that is empirically tractable, normatively principled, and practically actionable. The three-tier taxonomy---Confined Actors, Socially-Aware Interactors, and CPST-Integrated Agents---provides graduated governance that matches regulatory obligations to the actual depth of an entity's integration into human life. The composite assessment protocol, institutional design, and transition mechanisms developed in this paper provide the operational infrastructure necessary for regulatory implementation.

Getting the ontology right is not an academic exercise; it is the foundation on which all subsequent regulation will be built. We urge policymakers, standards bodies, and the research community to adopt the CPST-based classification framework and begin the work of standardization, empirical validation, and institutional design that coherent governance demands.


\end{document}